\documentclass[%
 reprint,
 amsmath,amssymb,
 aps,
]{revtex4-2}

\usepackage{graphicx}
\usepackage{dcolumn}
\usepackage{bm}

\usepackage{xr-hyper}
\begin{document}

\title{Orientational bistability and field-controlled switching \\of a superparamagnetic dimer}

\author{James R. N. Tett}
\affiliation{Physical and Theoretical Chemistry Laboratory, South Parks Rd, Oxford, OX1 3QZ United Kingdom}
\author{Finlay Johnston}%
\affiliation{Physical and Theoretical Chemistry Laboratory, South Parks Rd, Oxford, OX1 3QZ United Kingdom}
\author{Brennan Sprinkle}%
\affiliation{Department of Applied Mathematics and Statistics, Colorado School of Mines, Golden, CO, USA}
\author{Alice L. Thorneywork}%
\email{alice.thorneywork@chem.ox.ac.uk}
\affiliation{Physical and Theoretical Chemistry Laboratory, South Parks Rd, Oxford, OX1 3QZ United Kingdom}


\begin{abstract}
\noindent We study the orientational dynamics of superparamagnetic colloidal dimers that carry both an induced magnetic moment, proportional to the applied field, and an effective permanent moment. In a static, uniform magnetic field, dimers that are permanently fixed together hop between two preferred in-plane angles, developing a bimodal steady-state orientation distribution. When the same field is periodically reversed, we observe a sharp, field-controlled change in the dynamical response from small hopping events with $\Delta \theta\ll \pi$ to full $\Delta\theta \approx \pi$ rotations on each field flip. We show that both the static bistability and the switching bifurcation can be rationalised by a magnetic response in the dimer that consists of both a strong induced and weak body-fixed component. This leads to a complex orientational energy/potential landscape, with coupled roll-yaw rotations of the dimer responsible for the bistable dynamics. By combining the misorientation  between dimer axis and field, bifurcation field strength and short-time orientational variance, we determine the magnitude and orientation of the net permanent dipole, thereby characterising details of the internal magnetic structure of the particles via microscopy. 
\end{abstract}

\maketitle

\noindent Suspensions of superparamagnetic microparticles (SPMPs) find varied applications in nano- and biotechnology, serving as microactuators \cite{MICROACTUATORS_RANZONI,MAGNETIC_MICROROBOTS_TIERNO, MICROROBOTS_SPRINKLE}, biosensors  \cite{BIOSENSOR_RANZONI}, and transducers in single-molecule magnetic tweezing \cite{MAGNETIC_TWEEZERS_NEUMAN, MAGNETIC_TWEEZERS_CHOI_REVIEW}. More fundamentally, they  represent a versatile soft matter model system that has field-tunable, anisotropic interactions \cite{SIMULATING_DYNABEADS_TIERNO, MAGNETIC_ANISOTROPIC_INTERACTION_FAN, COLLOIDAL_SPIN_ICE_TIERNO}. This range of applications has motivated extensive effort to understand how magnetic field protocols control the structure and dynamics of particle assemblies, from effective pair interactions  \cite{MAGNETIC_ANISOTROPIC_INTERACTION_FAN} and field-driven chaining  \cite{MAGNETIC_CHAIN_SILVA} to collective organisation  \cite{EDGE_FLOW_MAGNETIC_COLLOIDS_NELSON} and transport in confined geometries \cite{MAGNETIC_COLLOIDS_CONFINEMENT_TIERNO}. In this context, magnetic dimers, i.e. an assembly of just two particles, are the simplest non-trivial structural component \cite{MAGNETIC_DIMERS_NING_WU}.

\noindent In many colloidal and biophysical settings, SPMPs are modelled as hard spheres with a purely induced dipole moment, that is proportional to, and instantaneously aligned with, the applied field \cite{REYNOLDS_AGGREGATION, MAGNETIC_DIMER_BOLTZMANN_INVERSION,SELF_ASSEMBLY_COLLOIDS_UNSTEADY_FIELDS_VICENTE}. In static, uniform fields, this induced-dipole picture is generally sufficient to rationalise forces and effective interaction potentials \cite{MAGNETIC_DIMER_BOLTZMANN_INVERSION, PARAMAGNETIC_COLLOIDS_THORNEYWORK, SUPERPARAMAGNETIC_TORQUE_DEKKER, COLLOIDAL_CLUSTERS_BISWAL}. Yet in many applications, the goal is to apply a magnetic torque to the bead; a process that is not possible in the above picture as it requires misalignment between the bead's net magnetic moment and the local field. As such, for rotating fields, the response is more subtle, involving 
magnetic anisotropy and/or internal degrees of freedom. Despite routine use of torque, for example, in magnetic tweezers,  
a full understanding of the bead-specific torque response remains elusive \cite{SUPERPARAMAGNETIC_TORQUE_DEKKER,MAGNETIC_TWEEZERS_CHOI_REVIEW, SLOW_RELAXATION_DYNAMICS_BISWAL}.

\begin{figure}
    \centering
    \includegraphics[width=\linewidth]{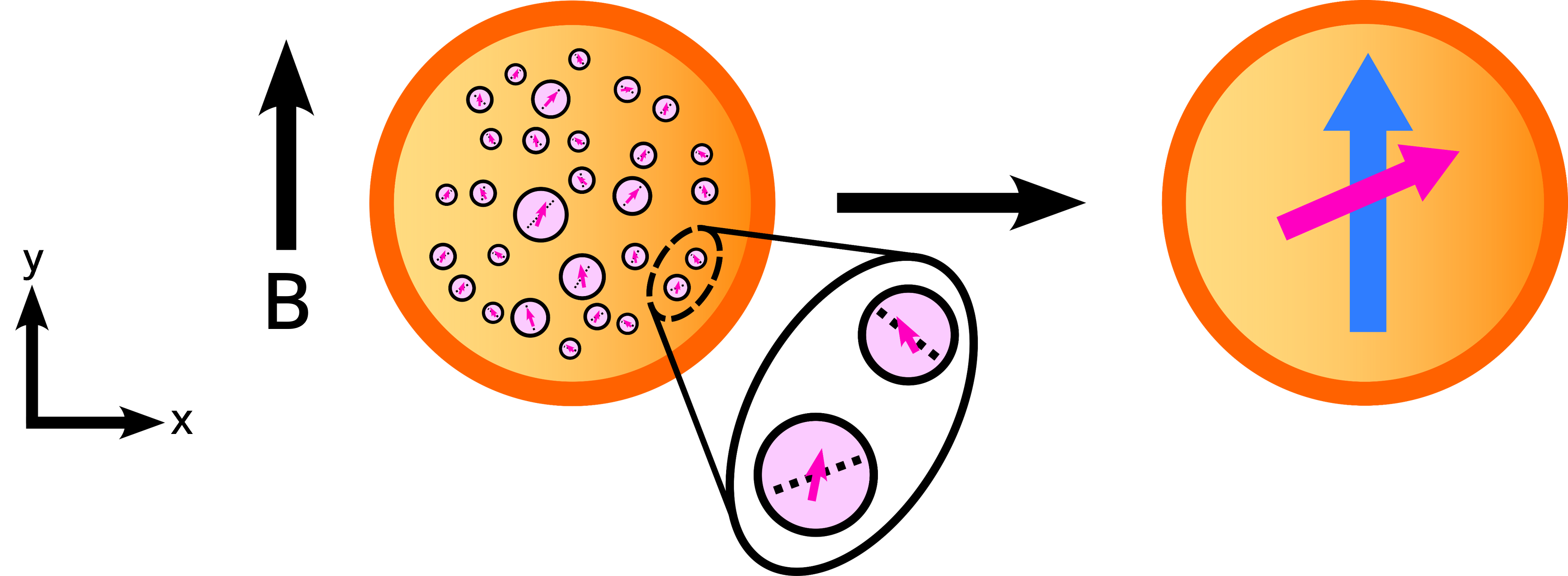}
    \caption{\textbf{Illustration of the internal structure and magnetic response of a superparamagnetic microparticle in an external field.}
    Random distribution of nanoparticles with net magnetic moments (lilac) within the polymer matrix results in regions of higher concentration. Associated with these is a magnetic easy axis (dashed line). In an external field, the collective response has two contributions: (i) a dominant induced moment, $\boldsymbol{\mu}_{\mathrm{ind}}$ (blue), aligned with the applied field $\mathbf{B}$, and (ii) a weaker body-fixed magnetic moment, represented as an effective permanent moment, $\boldsymbol{\mu}_{\mathrm{p}}$ (pink), fixed in the particle frame at a constant angle to the particle director. }
    \label{Fig:magneticstructure}
\end{figure}

\noindent Complexity around this question ultimately originates from the internal structure of SPMPs: a dilute dispersion of iron-oxide nanoparticles (NPs) embedded in a polymer matrix (Fig. \ref{Fig:magneticstructure}) \cite{PELECKY_RIEKE_MAGNETIC_PROPERTIES}. In an idealised scenario, on application of an external field the embedded nanoparticles respond independently: their magnetic moments are Boltzmann-distributed by competition between the Zeeman energy and thermal fluctuations, giving a net induced moment $\mu_{\mathrm{ind}} \propto B$ that is (on average) aligned with $\mathbf{B}$ and vanishes when the field is removed. Even at low nanoparticle volume fractions, however, if NPs are distributed randomly there is a finite probability that nearest neighbours will be close enough to interact and no longer behave as independent superspins \cite{DYNABEAD_CHARACTERISATION_FONNUM}. Such clusters can possess large effective anisotropy barriers and preferred magnetic axes \cite{CLUSTER_BARRIERS_OYARZUN}. At the coarse-grained level, this can give rise to the appearance of a weak, body-fixed moment superimposed on the dominant induced response; in effect, an apparent permanent moment(Fig. \ref{Fig:magneticstructure}). Experiments on commercial beads report both permanent contributions that can dominate low-frequency rotation  \cite{PERMANENT_DIPOLE_JANSSEN, PERMANENT_DIPOLE_PEASE} and an effective easy-axis anisotropy that becomes apparent at higher field strengths \cite{SUPERPARAMAGNETIC_TORQUE_DEKKER}, with details of the permanent contribution varying across measurements \cite{SUPERPARAMAGNETIC_TORQUE_DEKKER,SLOW_RELAXATION_DYNAMICS_BISWAL,PERMANENT_DIPOLE_PEASE}.  Many studies focus on high fields, however, where the induced moment dwarfs any residual contribution. In contrast, the weak-field regime, where even small body-fixed contributions can compete with the induced response to influence interaction potentials and dynamics, is less well explored. 

\noindent Here, we show that the orientational dynamics of superparamagnetic dimers in weak external fields reveals details of internal magnetic structure. We first explore behaviour in a static field for three different particle assemblies and confirm that the orientational dynamics of these different assemblies is consistent with coexisting induced and permanent moments in the weak-field regime. The field dependence of the effective interaction provides a scaling argument that distinguishes an effective permanent moment from anisotropic susceptibility effects. For two superparamagnetic particles rigidly fixed together we find that the dimer hops between two preferred orientations; a phenomenon that can be explained by considering the in-plane projection of a three-dimensional rotation of the net permanent moment. This establishes a complex two-dimensional potential landscape for dimer dynamics linked to coupled roll-yaw rotations of the dimer. Next, we consider how a dimer explores this complex landscape in response to flipping of the field direction. With increasing field magnitude, we find a sharp transition between two qualitatively different dynamics and again link this behaviour to details of the underlying landscape. For both static and flipping fields, we find quantitative agreement between experiment and theory, with these combined measurements allowing for the effective permanent dipole of individual dimers to be quantified from microscopy alone.

\section{Experimental methods}

\begin{figure*}
    \includegraphics[width=\textwidth]{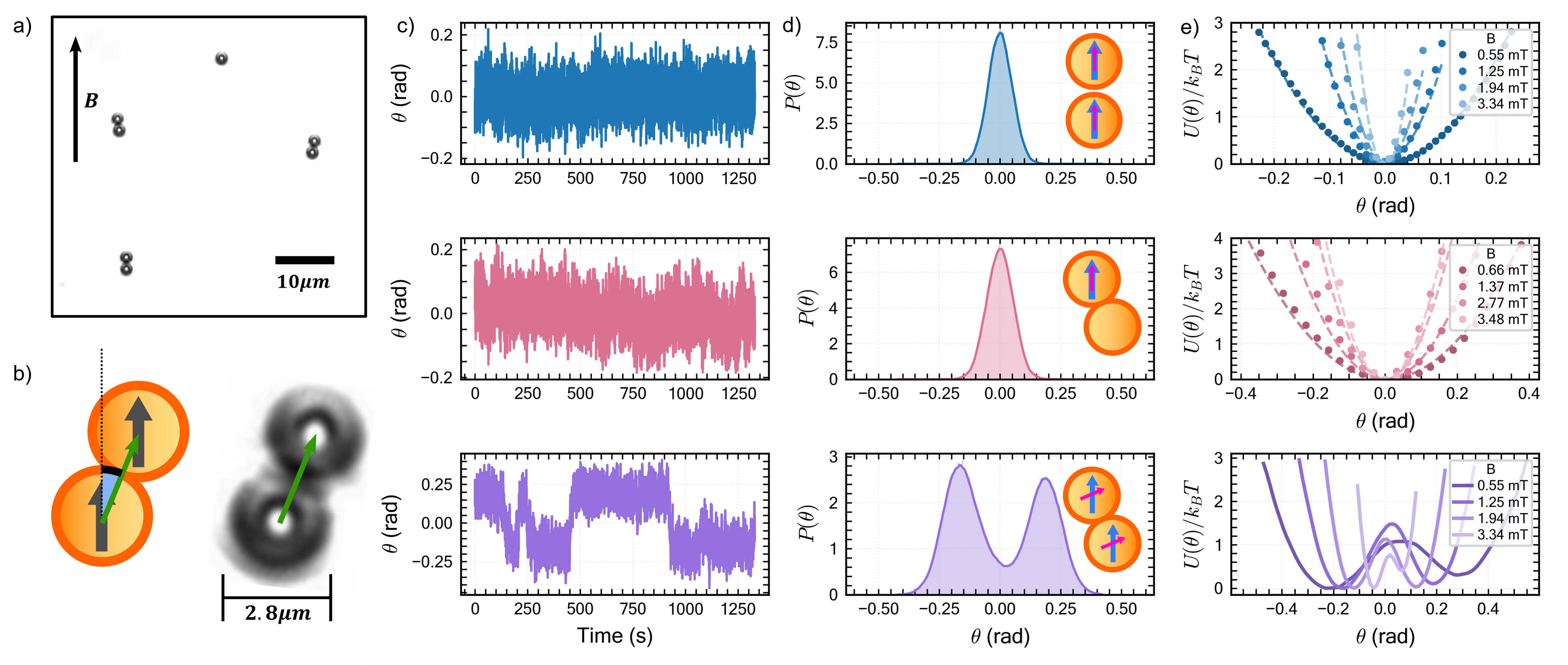}
    \caption{(a) Section of a typical experimental image showing magnetic dimers in an external field. The dimers shown are reversibly bonded such that removing the external field removes the attractive interaction. (b) Schematic of a magnetic-magnetic dimer showing the definition of the dimer axis and in-plane (yaw) angle with respect to the field. (c) Typical in-plane angle trajectories, $\theta(t)$ and (d) corresponding probability distribution of angle for: (top) two superparamagnetic particles interacting only via the field; (middle) a permanently bonded magnetic-non-magnetic dimer; (bottom) a permanently bonded magnetic-magnetic dimer. Data is taken at comparable field strengths ($\text{B}=1.25\text{mT}$ for magnetic-magnetic dimers and $\text{B}=1.37\text{mT}$ for magnetic-non-magnetic dimers). (e) Effective interaction potentials for (top) two superparamagnetic particles interacting only via the field; (middle) a permanently bonded magnetic-non-magnetic dimer; (bottom) a permanently bonded magnetic-magnetic dimer. (top) and (middle): Points show experimental data and dashed lines are fits to the dipole-dipole and dipole-field interaction potentials, Eqs.\ref{Eq: magmaginteraction} and \ref{Eq:Uhyb}, respectively. (bottom) solid lines show the experimental effective interaction potential.}
    \label{fig:experimental}
\end{figure*}

\noindent Our colloidal dimers were built from carboxylate-functionalised superparamagnetic beads of diameter $\sigma \simeq 3\mu\text{m}$ (Dynabeads M-270). These particles are reported to have an iron-oxide weight fraction of order $14\%$. \cite{MAGNETIC_SUSCEPTIBILITY_MEASUREMENTS} Suspensions were prepared at low volume fraction in deionised water and loaded into sealed glass sample cells of height $H \approx 200\mu\text{m}$. Rigid magnetic-magnetic dimers are found at low concentration in all samples and arise from irreversible surface interactions of two beads, giving a centre-centre separation $\ell \approx \sigma$. \cite{ERIC_WEEKS_DIMERS_IN_SAMPLE} Hybrid magnetic-non-magnetic dimers were prepared by mixing streptavidin-functionalised magnetic beads (Dynabeads M-280) with $1.9\mu\text{m}$ carboxylate-functionalised melamine-formaldehyde spheres (MF-COOH-S1000, MicroParticles GmbH).  Beads sedimented to the lower surface and the resulting out-of-plane fluctuations were small compared to $\sigma$ (gravitational height $h_g \approx 65\text{nm}\ll \sigma$), \cite{MAGNETIC_SUSCEPTIBILITY_MEASUREMENTS} so that the translational and orientational dynamics were quasi-two-dimensional.

\noindent A uniform, in-plane magnetic field $\mathbf{B}(t)$ was applied using a pair of coaxial Helmholtz coils driven by a programmable current source. We used both static fields, $\mathbf{B}=B\hat{\mathbf{y}}$, and square-wave protocols in which the field alternated between $\pm B\hat{\mathbf{y}}$ with period $T$. The field amplitude was calibrated as a function of coil current via a Hall probe and the field reversal time was small compared to the characteristic rotational relaxation time of the dimers ($\sim2\text{s}$).
Samples were imaged in bright-field using an inverted microscope equipped with a high-numerical-aperture objective and a Thorlabs CMOS CS135MU camera (Fig.~\ref{fig:experimental}a shows an example image). Time series were acquired at $25 \mathrm{fps}$ for up to $3\mathrm{hrs}$, sufficient to resolve intra-well Brownian fluctuations and the slower switching and flipping dynamics in static fields and field reversal. Particle coordinates were extracted using standard tracking methods \cite{CROCKER1996298}. From tracked bead positions, we obtained the dimer centre-of-mass and in-plane orientation $\theta(t)$, defined as the yaw angle between the dimer axis and $\hat{\mathbf{y}}$ (Fig. \ref{fig:experimental}b).

\noindent To quantify the rigidity of the bonded dimer assemblies, we tracked both constituent beads and computed the instantaneous bond length $\ell(t)$. The distribution of $\ell$ was narrowly peaked, with standard deviation comparable to the positional tracking uncertainty ($\sim 15\text{nm}, <1\%$ of a bead diameter), and showed no systematic dependence on field strength, indicating that bond-length fluctuations are minimised and that the particles are rigidly bound together. Over the same trajectories, dimer centres exhibited diffusive in-plane motion with $\langle \Delta r^2(t) \rangle \approx 4Dt$ (Supplemental Fig. S1),
consistent with free Brownian translation in the sedimentation plane. The mean squared angular displacements (MSADs, Supplemental Fig. S2)
were also proportional to $t$ at short times, consistent with angular diffusion within a potential well.

\section{Results and Discussion} 

\subsection{Dimer dynamics in static fields}

\noindent We first consider the orientational dynamics of individual assemblies in a static uniform field $\mathbf{B} = B\hat{\mathbf{y}}$. Fig. \ref{fig:experimental}c shows representative in-plane yaw trajectories $\theta(t)$ for three cases: (i) a pair of magnetic beads held together by induced dipole-dipole interactions (reversible bonding), (ii) a permanently linked magnetic-non-magnetic dimer, and (iii) a permanently linked magnetic-magnetic dimer. In cases (i) and (ii), $\theta(t)$ fluctuates around a single mean orientation. In contrast, rigid magnetic-magnetic dimers exhibit intermittent switching between two orientations. From long trajectories at fixed $B$, we construct a static-field yaw distribution $P(\theta)$ (Fig. \ref{fig:experimental}d). For pairs bound only by magnetic interactions (top) and the magnetic-non-magnetic dimers (middle), $P(\theta)$ is unimodal. By contrast, for rigid magnetic-magnetic dimers, $P(\theta)$ is bimodal across the measured field range. Rather than peaking at $\theta=0^{\circ}$ as expected for purely field-aligned induced dipoles, the distribution develops two preferred orientations at $\pm \theta_0(B)$, where $ \theta_0(B)$ is on the order of $10-20^\circ$,  separated by a lower-probability region near $\theta=0$.

\noindent To visualise the effective angular landscape, we construct $U(\theta) = -k_B T \ln \left( P(\theta) \right) + \text{const.}$ from the measured angular distribution and plot $U(\theta)$ for each dimer type at several field strengths (Fig. \ref{fig:experimental}e). For the dimer interacting only via the magnetic field (top), experimental data (circles) are in good agreement with the expected form of the landscape arising from an induced dipole-dipole interaction between two moments aligned with $\mathbf{B}$ (lines) given by:
\begin{equation}
    U_{\mathrm{ind}}(\theta) = -\boldsymbol{\mu}_{\mathrm{ind}, 1} \cdot \mathbf{T}_{12}(\theta) \cdot \boldsymbol{\mu}_{\mathrm{ind},2}
    \label{Eq: magmaginteraction}
\end{equation}

\noindent where $\mathbf{T}_{12} = \frac{\mu_0}{4 \pi r^3} \left( \mathbf{I} - 3 \hat{\mathbf{r}} \hat{\mathbf{r}} \right)$ is the dipolar coupling tensor and $\boldsymbol{\mu}_{\mathrm{ind, i}} = \chi B \hat{\mathbf{y}}$ in the linear-response regime. To leading order, an induced dipole-only picture is sufficient to describe the behaviour. For the magnetic-non-magnetic dimer, a purely induced moment that is strictly collinear with $\mathbf{B}$ would produce no torque in a uniform field, and so there should not be a preferred angle for the dimer. The observed preferred yaw angle requires a body-fixed contribution, consistent with the particle carrying a weak residual or permanent moment on experimental timescales. 

\noindent We quantify this moment by fitting the magnetic-non-magnetic landscapes (Fig. 2(e), middle) to the minimal permanent-dipole form:
\begin{equation}
    U_{\mathrm{hyb}}(\theta) = - \mu_{\mathrm{p}} B \cos(\theta - \alpha)
    \label{Eq:Uhyb}
\end{equation}

\noindent where $\mu_{\mathrm{p}}$ is an effective body-fixed moment (i.e. a coarse-grained odd-in-$B$ contribution fixed in the particle frame) and $\alpha$ sets its orientation projected onto the plane of the sample. This single-harmonic description captures both the unimodal shape of $P(\theta)$ and is consistent with the fact that (i) dimers are observed to rotate by $180^\circ$ upon field inversion (Supplemental Fig. S5)
and (ii) have an interaction strength proportional to $B$, rather than $B^2$ (Fig. S6).
These features distinguish the odd-in-$B$ contribution from induced magnetisation, which, even in the presence of anisotropic susceptibility, should remain even-in-$B$ and, in the linear-response regime, should scale proportional to $B^2$. From the fits to data for the magnetic-nonmagnetic dimer in Fig.~\ref{fig:experimental}(e), we obtain a linear scaling of interaction strength with field and $\mu_{\mathrm{p}} \sim (0.5\text{-}5) \times 10^{-16} \mathrm{A m^2}$ for individual particles. This is consistent with previously reported permanent-moment magnitudes for commercial superparamagnetic beads \cite{PERMANENT_DIPOLE_PEASE,HYDRODYNAMICS_COUGHLAN}.

For rigid magnetic-magnetic dimers, the bimodal $P(\theta)$ again indicates that an additional ingredient is at play beyond the induced interaction. While a natural first hypothesis is that the two peaks reflect two independent body-fixed moments, one in each bead, any combination of two permanent moments in the plane of the sample can be expressed in terms of a single effective in-plane dipolar torque, yielding only a single minimum. Rationalising the bimodal distribution, therefore, requires a fuller consideration of the orientation of the permanent moment. 

\begin{figure}
\includegraphics[width=\linewidth]{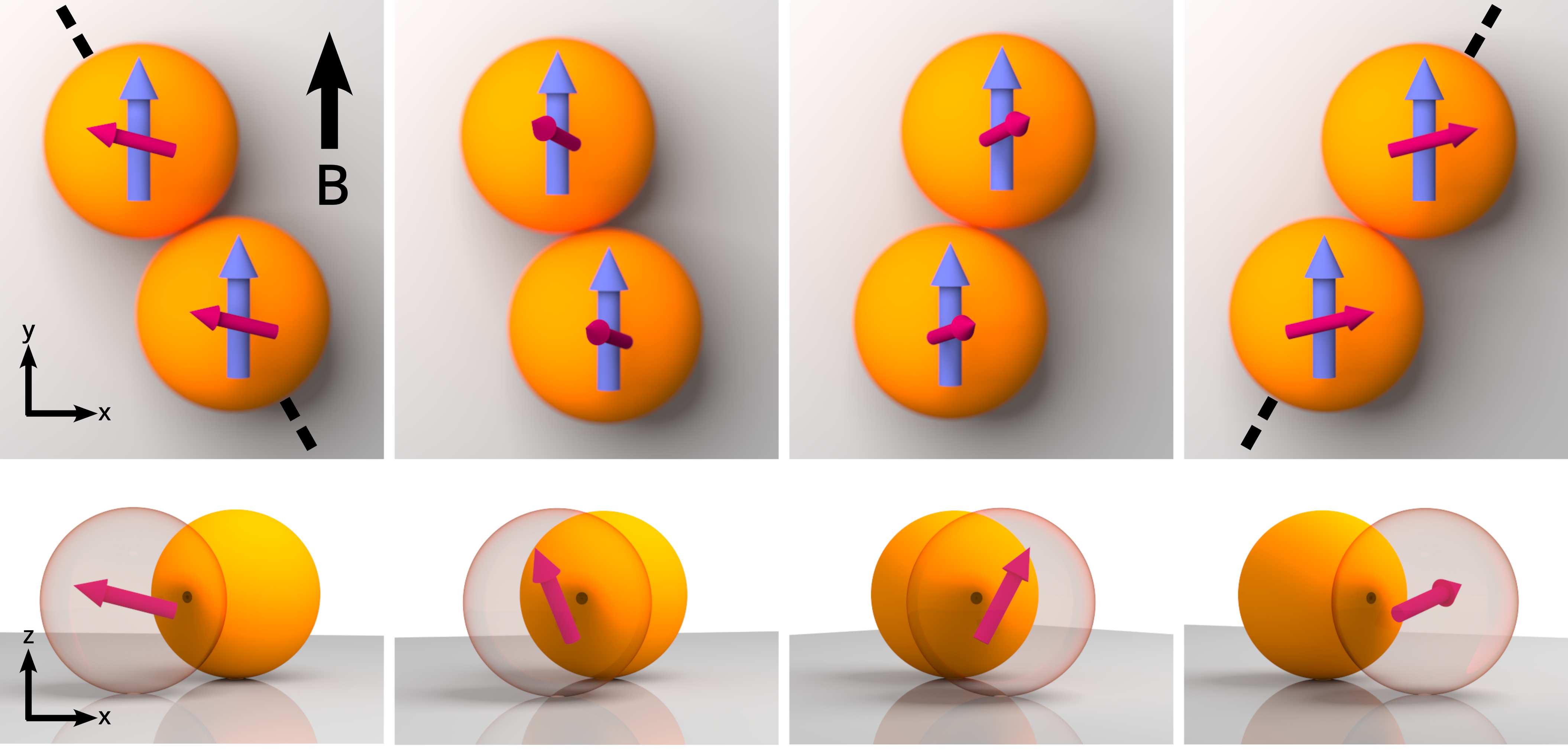}
    \caption{Schematic sequence (left to right) showing the continuous reorientation of a rigid dimer through coupled roll and yaw rotations when viewed from above (upper row) and in the sedimentation plane (lower row). Roll states are shown via the permanent moment orientation. Only one permanent moment is shown for clarity (lower row).}
    
    \label{Fig:rollyaw+landscape}
\end{figure}

\begin{figure*}
    \centering
    \includegraphics[width=\linewidth]{Main_Text_Figures/Figure_4/figure_4_v8.png}
    \caption{(a) Example 2D roll-yaw energy landscape $U(\theta,\rho;B)$ for a rigid dimer in a static field applied along $\theta=0\text{ rad}$. Circles mark the positions of global minima (permanent moments broadly aligned with the external field). Stars show the position of local minima (permanent moments broadly misaligned with the external field). Representative dimer configurations are shown for minima.  (b) 1D cuts through the landscape in (a) corresponding to an in-plane, yaw only rotation (black dashed line) and a coupled roll-yaw path (orange dashed line).  (c) Variation in the misalignment between dimer and field direction as a function of applied external field, $B$, for multiple different dimers. Points show experimental data and lines show fits according to Eq.~\eqref{eq: stationarity condition}. }
    \label{fig: U(theta) and theta_min vs. B}
\end{figure*}

While the dimer itself is confined to diffuse in a 2D plane, the effective permanent moment could have any three-dimensional orientation within the particle.
As the dimer diffuses, it is therefore possible for the permanent contribution to move out of plane due to a roll about the dimer axis (see Fig.~\ref{Fig:rollyaw+landscape}(a), where we denote the roll angle as $\rho$). This roll is not directly observable in our images due to the particle symmetry. For most orientations of the permanent moment, a change in $\rho$ changes the projection of the permanent moment in the plane of the sample, and thus the preferred in-plane, or yaw, angle. If the dimer itself had unrestricted 3D motion, it would precess about the external field, and the hopping between two states that we observe arises from the physical confinement by gravity.

\subsubsection{Roll-yaw landscapes for rigid magnetic dimers}

To understand the bimodal $P(\theta)$ distribution, we hypothesize that it arises as a marginalised projection of a coupled $(\theta,\rho)$ landscape, and so explore the full roll-yaw energy landscape $U(\theta,\rho)$ of the dimer in detail.
\noindent To construct the landscape, we modify Eq.~\ref{Eq: magmaginteraction} to include the interaction between the permanent moment and external-field term as:

\begin{equation}
    U(\theta,\rho) = -\boldsymbol{\mu}_{\mathrm{p}, \mathrm{net}}(\theta,\rho) \cdot \mathbf{B} - \boldsymbol{\mu}_{\mathrm{ind},1} \cdot \mathbf{T}_{12}(\theta,\rho) \cdot \boldsymbol{\mu}_{\mathrm{ind},2}
    \label{eq: U_2D_main}
\end{equation}
where $\mu_{\mathrm{p}, \mathrm{net}}$ is the magnitude of the dimer's net permanent moment. The induced and permanent terms are even and odd in $B$ respectively and mutual induction and permanent-induced cross terms produce only small corrections in our geometry (see Supplemental Section S4).

Eq. \ref{eq: U_2D_main} predicts a family of energy landscapes. The features of these depend on the relative values of induced and permanent magnetic interactions, as set by the external field strength and the magnitude and orientation of the permanent moments within each bead. Fig.~\ref{fig: U(theta) and theta_min vs. B}(a) shows a typical example of such a landscape with experimentally reasonable values of  $B=3\text{mT}$ and $\left[|\boldsymbol{\mu}_{p,net}|, \text{arg}(\boldsymbol{\mu}_{p,net}) \right]= \left[ 6.6\times10^{-16} \text{Am}^2, 45^\circ \right]$. The landscape shows multiple local minima related by a glide symmetry $U(\theta,\rho) = U(-\theta,\rho+\pi)$ and we focus first on the global minima (circles in Fig.~\ref{fig: U(theta) and theta_min vs. B}(a)). Starting from one such minima, a cut across the landscape that varies in only the yaw angle, $\theta$, (black dashed line) clearly shows only one accessible minima (Fig.~\ref{fig: U(theta) and theta_min vs. B}(b)). In contrast, moving diagonally across the landscape (orange) – achieved by varying both the in-plane yaw angle, $\theta$ while rotating around the dimer axis by angle $\rho$ – we find a pathway connecting two minima with a barrier on the order of $4 k_B\text{T}$, i.e., an accessible bistable potential. This supports our physical interpretation of the dimer hopping as arising from coupled roll-yaw motion. Interestingly, in addition to the series of global minima, which correspond to permanent moments with orientation broadly aligned with the field, a second set of minima exist (diamonds in Fig.~\ref{fig: U(theta) and theta_min vs. B}(a)) that correspond to a mismatch between external field and permanent moment orientations. These are locked out of the global minima by the enormous energy barriers between the two branches, but would lead to equivalent hopping behaviour. 

\subsubsection{Misorientation Angle}
\noindent Next, we seek to explain quantitatively  the characteristic behaviour of the bimodal $U(\theta)$ as a function of field strength in Fig. \ref{fig:experimental}(e), namely, the variation in minima position with field strength. From the experimental $P(\theta)$ distributions, we extract the position of the two roll-related wells, $\pm \theta_0(B)$ with respect to the applied external field. These are plotted as a function of field strength for five different dimers in Fig.~\ref{fig: U(theta) and theta_min vs. B}(c). 
For high fields, we find that $\theta_0(B) \propto 1/B$ but the data deviates from this at low fields, where the misorientation angle is below a $1/B$ trend. The absolute values of the misorientation angle are also specific to a particular dimer. 

\noindent To compare our experimental data quantitatively to the roll-yaw landscapes, we obtain an effective 1D yaw potential that now includes the effect of the permanent moment as:
\begin{equation}
    U_{\mathrm{eff}}(\theta;B) = -\kappa B^2 \cos(2\theta) - \mu_{\mathrm{p}, \mathrm{net}} B \cos(\theta - \alpha),
    \label{eq: U_1D_main}
\end{equation}
\noindent where $\kappa$ is the induced dipole-dipole interaction stiffness and $\alpha$ is the in-plane misorientation of the net permanent moment relative to the dimer axis (see Supplemental Section S7).
Applying the glide symmetry generates partner minima on the second roll branch, recovering the paired minima observed experimentally (Fig. \ref{fig:experimental}e). We note that equation \eqref{eq: U_1D_main} has the same mathematical structure as a Stoner-Wohlfarth energy \cite{Stoner_Wohlfarth_1948}, with the induced dipole-dipole interaction providing the effective two-fold anisotropy and the permanent moment providing the Zeeman-like tilt. 

\noindent On differentiation, Eq.~\ref{eq: U_1D_main} yields an explicit expression for the positions of potential minima as a function of applied field as
\begin{equation}
    2 \kappa B \sin(2\theta_{\min}) + \mu_{\mathrm{p,net}} \sin(\theta_{\min} - \alpha) = 0.
    \label{eq: stationarity condition}
\end{equation}

\noindent In principle, Eq.~\ref{eq: stationarity condition} contains three unknowns: the induced dipole-dipole interaction stiffness, $\kappa$, the in-plane misorientation of the net permanent moment relative to the dimer $\alpha$ and the net permanent moment, $\mu_{p,net}$. The stiffness, $\kappa$, can, however, be determined from short-time intra-well fluctuations $\langle \Delta \theta^2 \rangle$ (see Supplemental Section S6).
As such, fitting Eq. \eqref{eq: stationarity condition} to individual dimer data determines two dimer-specific parameters $\mu_{\mathrm{p,net}}$ and $\alpha$, on a dimer-by-dimer basis. Measurement of intra-well fluctuations for rigid magnetic-magnetic dimers, yields a magnetic susceptibility of $(1-10)\times10^{-12} \text{Am}^2 \text{T}^{-1}$, consistent with literature measurements  \cite{MAGNETIC_SUSCEPTIBILITY_MEASUREMENTS}. Calculated magnetic moments are on the order $(0.5-3.0) \times 10^{-16} \text{Am}^2$, likewise in line with dimer literature measurements  \cite{DIMER_ESTIMATION_LITERATURE,PERMANENT_DIPOLE_PEASE}, with misorientation angle $| \alpha | \sim (0.3-2.2)\text{rad}$. It should be noted that small misorientation angles ($\alpha$) are difficult to determine with this method, as dimer axis measurements will be masked by noise.

\subsection{Dimer dynamics in flipping fields}


\begin{figure*}
    \centering
    \includegraphics[width=\linewidth]{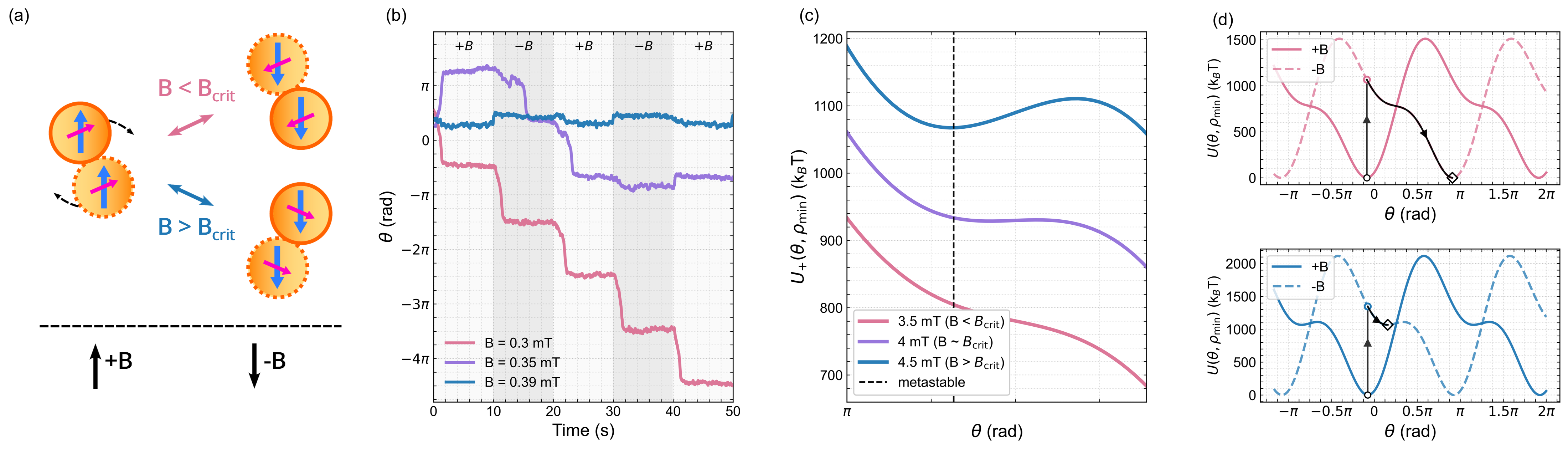}
    \caption{(a) Schematic showing the two dimer responses to field flips. Below the bifurcation field, the dimer undergoes full $\pi$ reorientations (full flips) whilst above the bifurcation field, the dimer undergoes small reorientations (small hops). (b) Experimental angular trajectories for the same rigid dimer above (blue), around (purple) and below (orange) the critical field, $B_{\mathrm{crit}}$. (c) 1D cut-throughs of an example roll-yaw landscape, $U(\theta, \rho=\rho_{\text{min}})$, for increasing field magnitude. Note the development of a metastable minimum in the blue potential as the field passes through the bifurcation. (d) Relaxation under field reversal: 1D effective yaw potentials $U(\theta, \rho=\rho_{\text{min}})$ before and after field reversal, $+B \rightarrow -B$, for field magnitudes below (top) and above (bottom) the bifurcation field. Below $B_{\text{crit}}$, reversal drives relaxation by a large $\sim \pi$ reorientation, whereas above $B_{\mathrm{crit}}$ the appearance of a metastable minimum leads to a small-angle hop.}
    \label{fig: Field Flip Response}
\end{figure*}

\noindent Having characterised how the presence of a permanent moment influences the dynamics of rigid dimers in a static field, we now consider dimer dynamics in a flipping field. In particular, we apply a square-wave protocol, in which the field alternates between $\pm B \hat{\mathbf{y}}$ with a period much longer than the rotational relaxation time for the dimer. For purely induced dipole interactions, flipping the field by $\pi$ leads to a corresponding switch in the direction of the induced dipole within the bead, and thus should not lead to rotation. In contrast, for the permanent moment, flipping the field direction will generate a more or less favourable situation, depending on whether the permanent moment is initially net aligned or misaligned with the field. 

Fig. \ref{fig: Field Flip Response}(a) and (b) show the possible relaxation responses observed for rigid dimers in response to a field flip. At lower $|B|$ (pink), the dimer rotates by $\sim 180^\circ$ on each field reversal, but does not show the small-angle hopping observed in a static field. At higher $|B|$ (blue), the dimer executes small-angle hops between the two static-field wells at $\pm \theta_0(B)$. In both cases, we observe both clockwise and anticlockwise rotations, in line with the sequence of minima in the energy landscape. Example experimental trajectories, showing the angle of the dimer as a function of time, can be seen in Fig. \ref{fig: Field Flip Response}(b). At intermediate fields (purple), we find a mixed scenario with trajectories exhibiting both small-angle hops and full rotation by $180^\circ$. Remarkably, the transition in behaviour between full rotations and small hops occurs over a very small range of field strengths.

\noindent These different trajectories can also be understood in terms of the roll-yaw landscape $U(\theta,\rho;B)$ (Eq. \eqref{eq: U_2D_main}). Upon field reversal $B \rightarrow -B$, the instantaneous configuration of the dimer is mapped to a high-energy region of $U(\theta,\rho;-B)$ (Fig. \ref{fig: Field Flip Response}(d)), and the dimer relaxes toward a new stable minimum. For low field strengths, i.e., relatively weak induced moment interactions, 
the permanent moment (odd-in-$B$ term) dominates, leading to large $\sim\pi$ reorientation. As B increases, the strength of the induced moment grows, and the post-flip landscape develops metastable minima in which the induced moments remain well-aligned while the permanent moment is misaligned (Fig. \ref{fig: Field Flip Response}(c)). Relaxation therefore proceeds via small hops between nearby wells. 

\begin{figure*}[t]
\centering
\begin{minipage}[c]{0.69\textwidth}
    \centering
    \includegraphics[width=0.95\linewidth]{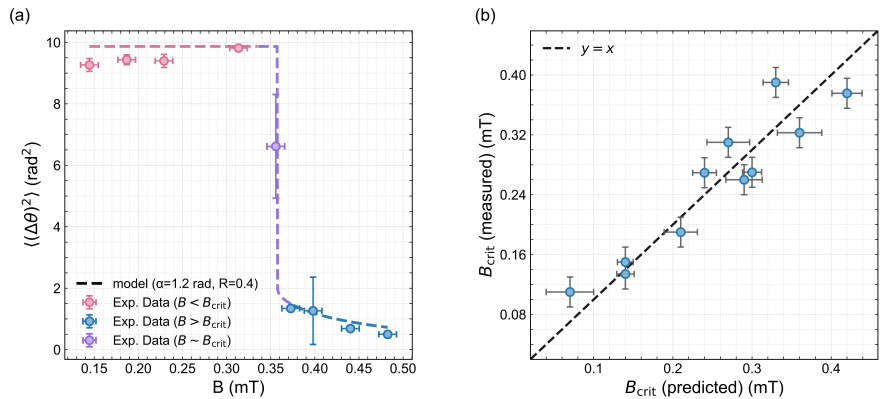}
\end{minipage}\hfill
\begin{minipage}[c]{0.30\textwidth}
    \caption{(a) Mean squared angular step size, $\langle (\Delta \theta)^2 \rangle$, as a function of $B$, demonstrating the bifurcation in switching behaviour. Error bars denote the standard error of the mean over switching events at each field ($n \approx 10$-$20$ events). (b) Comparison of the experimentally measured $B_{\text{crit}}$ and prediction from Eq. \eqref{eq: B_crit} for multiple dimers. Error bars indicate standard errors in extracted experimental $B_{\text{crit}}$ and in the predicted value obtained by propagating uncertainties in fitted parameters. The variation in $B_{\text{crit}}$ can be attributed to heterogeneity in permanent moment strengths and orientations across dimers.}
    \label{fig:bifurcation plot}
\end{minipage}
\end{figure*}

The two different regimes can be characterised in more detail via the mean squared step size upon field inversion for a trajectory,  $\langle (\Delta \theta)^2 \rangle$, as shown in Fig \ref{fig:bifurcation plot}(a). This exhibits a sharp crossover from full flipping of the dimer to small hops as $|B|$ is increased, with the mixed trajectory intermediate between these cases.


\noindent The bifurcation in field reversal response can be captured quantitatively by a spinodal condition in the effective yaw landscape: at a critical field magnitude $|B|=B_{\mathrm{crit}}$, a metastable minimum on a given branch disappears. This is defined by:

\begin{equation}
    \frac{\partial U_{\mathrm{eff}}}{\partial \theta} = 0, \qquad \frac{\partial^2U_{\mathrm{eff}}}{\partial \theta^2} = 0
    \label{eq: spinodal condition}
\end{equation}

\noindent For Eq. \eqref{eq: U_1D_main}, this yields:

\begin{equation}
    B_{\mathrm{crit}} = \frac{\mu_{\mathrm{p,net}}}{2 \kappa f(\alpha)}
    \label{eq: B_crit}
\end{equation}

\noindent with the angular shape factor:

\begin{equation}
    f(\alpha) = \Bigl(|\cos(\alpha)|^{2/3} + |\sin(\alpha)|^{2/3}\Bigr)^{3/2}
    \label{eq: angular shape factor (astroid)}
\end{equation}

\noindent Eq. \eqref{eq: angular shape factor (astroid)} has the astroid form familiar to the Stoner-Wohlfarth switching boundary  \cite{Stoner_Wohlfarth_1948,THIAVILLE19985}, with the role of an easy-axis anisotropy replaced here by the induced dipole-dipole term $\propto B^2$ (Eq. \eqref{eq: U_1D_main}). Using $\mu_{\mathrm{p,net}}$ and $\alpha$ obtained from the static well positions and $\kappa$ from intra-well fluctuations, Eq. \eqref{eq: B_crit} predicts $B_{\mathrm{crit}}$ for each dimer without additional fitting, in agreement with the crossover observed in $\langle (\Delta \theta)^2 \rangle$. This is shown as the dashed line in Fig. \ref{fig:bifurcation plot}(b). Comparison between predicted and observed $B_{\mathrm{crit}}$ is good within error across dimers, with critical fields ranging from $(0.1-1.0)\mathrm{mT}$ (Fig. \ref{fig:bifurcation plot}(b)). 
\newline
\newline
\noindent \textit{Single-dimer magnetometry: } Taken together, the static misorientation $\theta_0(B)$, high-field intra-well stiffness $\kappa$, and crossover field $B_{\mathrm{crit}}$ provide a practical route to estimating the net permanent dipole of an individual dimer from microscopy. First, at sufficiently large $B$, the local curvature of $U_{\mathrm{eff}}(\theta;B)$ is dominated by the induced term. By approximating the dynamics about a single minimum as an overdamped Ornstein-Uhlenbeck process, the short-time variance $\langle \Delta \theta^2 \rangle$ yields the induced stiffness $\kappa$ for that dimer. Second, the measured well positions $\theta_0(B)$ at one or more intermediate fields constrain $\mu_{\mathrm{p,net}}/\kappa$ and the in-plane angle $\alpha$ via Eq. \eqref{eq: stationarity condition}. Finally, the independently determined $B_{\mathrm{crit}}$ provides an additional constraint through Eq. \eqref{eq: B_crit}. In this way, combined static and flipping-field measurements act as a single-object rotational magnetometry protocol for anisotropic superparamagnetic dimers, without requiring separate bulk magnetometry. Whilst our method fits the $B_{\mathrm{crit}}$, a limitation of this method is its bias towards large misorientation angles, as $\theta_0(B)$ vs. $B$ can be measured more accurately. By applying this protocol, we observe permanent moments of $(0.5-3.0) \times 10^{-16} \text{Am}^2 $ (consistent with literature values  \cite{DIMER_ESTIMATION_LITERATURE}) with characteristic misorientation angles ranging widely, from $(0.3-2.2)\mathrm{rad}$ for our measured dimers.
\vspace{-1.0em}
\section{Conclusions}
\vspace{-0.5em}
\noindent Our results show that weak effective permanent moments in nominally superparamagnetic microparticles can have measurable consequences for the orientational dynamics of even the simplest assembly, a rigid dimer. A minimal roll-yaw landscape comprising an even-in-$B$ induced term and an odd-in-$B$ permanent-dipole tilt quantitatively accounts for (i) the odd-in-$B$ preferred orientation of magnetic-non-magnetic dimers, (ii) the bistable static yaw landscape of rigid magnetic-magnetic dimers, and (iii) the sharp crossover in switching behaviour under periodic field reversal. The same framework suggests a practical single-object protocol to extract the magnitude and in-plane orientation of the dimer's net permanent moment from direct imaging.

\noindent More broadly, these results identify rigid magnetic dimers as a minimal soft-matter system in which internal magnetic asymmetries are converted into multistable orientational dynamics and field-controlled switching pathways. This behaviour is relevant to the interpretation of torque-generating bead assays and to the design of simple magnetic microactuators, especially in weak-field regimes. More generally, this work shows that the microscopic magnetic heterogeneity of superparamagnetic microparticles leaves a clear dynamical fingerprint at the level of individual colloidal assemblies. \newline
\section*{Author contributions}
\vspace{-0.5em}
\noindent J.R.N.T.: conceptualisation, methodology, investigation, data curation, formal analysis, visualisation, writing - original draft. F.J.: investigation, data curation, writing - review \& editing. B.S.: formal analysis, visualisation, writing - review \& editing. A.L.T.: conceptualisation, methodology, supervision, writing - review \& editing.
\vspace{-1.0em}
\section*{Conflicts of interest}
\vspace{-0.5em}
\noindent There are no conflicts to declare.
\vspace{-1.5em}
\section*{Data availability}
\vspace{-0.5em}

\noindent All data needed to evaluate the conclusions in the paper are present in the paper and/or the Electronic Supplementary Information (ESI). Raw microscopy videos, particle-tracking data, and additional analysis files are available from the corresponding author upon reasonable request.
\vspace{-1.5em}
\section*{Acknowledgements}
\vspace{-0.5em}
We wish to acknowledge fruitful discussions with Kurt Andresen, Cheuk Kit Ngai, Anna Drummond Young and Eleanor Mackay. J.R.N.T. and A.L.T. acknowledge funding from EPSRC (EP/X02492X/1). A.L.T. acknowledges funding from a Royal Society University Research Fellowship (URF/R1/211033).

\nocite{*}

\bibliography{bibliography}

\end{document}